\documentclass{emulateapj}

\usepackage{color}
\usepackage{graphicx}
\usepackage{natbib}
\slugcomment{Received 2017 September 8; revised 2018 January 8; accepted 2018 January 26}

\shorttitle{VLBA Imaging of Double-Peaked Narrow-Line AGNs}
\shortauthors{Liu, Lazio, Shen, Strauss}

\newcommand{\mjybm}{\mbox{mJy~beam${}^{-1}$}}

\newcommand{\foiii}{\mbox{[\ion{O}{3}]}}
\newcommand{\OIIIc}{\mbox{[\ion{O}{3}]}\,$\lambda\lambda$4959,5007}
\newcommand{\Msun}{\ensuremath{M_{\sun}}}

\begin{document}

\title{Very Long Baseline Array Imaging of Type-2 Seyferts with Double-Peaked Narrow Emission Lines: Searches for Sub-kpc Dual AGNs and Jet-Powered Outflows\altaffilmark{*}}

\altaffiltext{*}{Based, in part, on observations made with the Very Long Baseline Array, obtained at the Long Baseline Observatory. The Long Baseline Observatory is a facility of the National Science Foundation operated under cooperative agreement by Associated Universities, Inc. }

\author{Xin Liu\altaffilmark{1,2,6}, T.~Joseph~W.~Lazio\altaffilmark{3,7}, Yue Shen\altaffilmark{1,2,5}, and Michael A.~Strauss\altaffilmark{4}}

\altaffiltext{1}{Department of Astronomy, University of Illinois at Urbana-Champaign, Urbana, IL 61801, USA}

\altaffiltext{2}{National Center for Supercomputing Applications, University of Illinois at Urbana-Champaign, 605 East Springfield Avenue, Champaign, IL 61820, USA}

\altaffiltext{3}{Jet Propulsion Laboratory, California Institute of Technology, M/S~138-308, 4800 Oak Grove Drive, Pasadena, CA  91109, USA}

\altaffiltext{4}{Department of Astrophysical Sciences, Princeton University, Peyton Hall, Ivy Lane, Princeton, NJ  08544, USA}

\altaffiltext{5}{Alfred P. Sloan Foundation Fellow}

\altaffiltext{6}{Email: xinliuxl@illinois.edu}

\altaffiltext{7}{Email: Joseph.Lazio@jpl.nasa.gov}

\begin{abstract}
This paper presents Very Long Baseline Array (VLBA) observations of 13 double-peaked \foiii\ emission-line type-2 active galactic nuclei (AGNs) at redshifts $0.06<z<0.41$ (with a median redshift of $z\sim0.15$) identified in the Sloan Digital Sky Survey. Such double-peaked emission-line objects may result from jets or outflows from the central engine or from a dual \hbox{AGN}. The VLBA provides an angular resolution of $\lesssim$10~pc at the distance of many of these galaxies, sufficient to resolve the radio emission from extremely close dual AGNs and to contribute to understanding the origin of double-peaked \foiii\ emission lines. Of the 13 galaxies observed at~3.6~cm (8.4~GHz), we detect six at a 1$\sigma$ sensitivity level of $\sim$0.15 mJy beam$^{-1}$, two of which show clear jet structures on scales ranging from a few milliarcseconds to tens of milliarcseconds (corresponding to a few pc to tens of pc at a median redshift of 0.15). We suggest that radio-loud, double-peaked emission-line type-2 AGNs may be indicative of jet produced structures, but a larger sample of double-peaked \foiii\ AGNs with high angular resolution radio observations will be required to confirm this suggestion. 
\end{abstract}

\keywords{black hole physics -- galaxies: active -- galaxies: interactions --- galaxies: nuclei --- galaxies: Seyfert -- radio continuum: galaxies -- quasars: general}

\section{Introduction}\label{sec:intro}

\subsection{Significance of Binary Supermassive Black Holes}

It is now well established that most, if not all, large galaxies in the local universe have supermassive black holes (SMBHs; $> 10^8\,\Msun$) in their nuclei \citep[e.g.,][]{kormendy95,ff05}. Further, the merger of galaxies is considered to be an integral aspect of galaxy assembly and evolution, with large galaxies in the local universe potentially having undergone multiple mergers during the course of their evolution. A robust prediction of the galaxy merging process is that the merger product will contain two SMBHs, which will sink to the center of the product via dynamical friction on time scales of order $10^8$ yr.  Models of this merger process not only predict the formation of a binary\footnote{We distinguish between binary SMBHs and dual SMBHs or dual active galactic nuclei (AGNs). The former are systems for which their mutual gravitational interactions are dominant, while the latter are galaxies or merger products that contain two SMBHs, but for which the separation between the two SMBHs is so large that their motions are dominated by the gravitational potential of the host. } \hbox{SMBH} \citep[e.g.,][]{begelman80,milosavljevic01,yu02,DEGN}, but also claim success in being able to replicate a variety of other aspects of galaxy or quasar properties, including the $M_{\mathrm{BH}}$-$\sigma$ relation, the quasar luminosity function, the central brightness of galaxies, and the bending or apparent precession of radio jets \citep[e.g.,][]{kauffmann00,volonteri03,wyithe03,liufk04,hopkins08, Kormendy2009a,shen09,KormendyHo2013}.

The kinematics of the two SMBHs at the center of the merger product begins to be dominated by their mutual gravitational interaction, rather than by the gravitational potential of the host, when their separation is \citep{begelman80,volonteri03}
\begin{equation}
r_b = \frac{G(m_1 + m_2)}{2\sigma^2}
    \sim 10\,\mathrm{pc}\,\frac{m_1 +
    m_2}{10^8\,\Msun}\left(\frac{\sigma}{150\,\mathrm{km}\,\mathrm{s}^{-1}}\right)^{-2},
\label{eqn:binary}
\end{equation}
for SMBHs of mass~$m_1$ and~$m_2$ located within a merger product with a central velocity dispersion~$\sigma$. Further dynamical friction by stars in the central region, and possibly gas interactions, cause the binary to harden.  For a time, it was thought that the binary separation would cease to shrink at a separation of order 1 pc as stellar interactions become less effective \citep[the ``last parsec problem,''][]{begelman80,Quinlan1996,milosavljevic01,yu02}. Considerable recent attention has focused on interactions between the SMBH binary and the surrounding population of stars, particularly in light of the fact that the stellar population is likely to have an asymmetric spatial or velocity distribution as a result of the merger process itself \citep[e.g.,][]{yu02,berczik06,Mayer2007,Lodato2009,Sesana2010,khan11,Preto2011}. While the results are not yet conclusive, it appears plausible that such interactions would cause the binary to continue to harden to sub-milliparsec separations within the age of the universe, at which point its separation will shrink inexorably due to the emission of gravitational waves (GWs). Current or future pulsar timing arrays should be able to detect the GW emission from the ensemble of individual binary SMBHs at frequencies of order $10^{-9}$~Hz \citep{hobbs10,Arzoumanian2016,Babak2016}, while the last moments of in-spiral should produce GWs with frequencies of order $10^{-6}$~Hz, which would be detectable by future space interferometers \citep{Amaro-Seoane2013,Danzmann2017}.  Determining the cosmological density of galaxies containing a pair of SMBHs therefore provides constraints on the rate at which galaxies undergo mergers and the late stages of the merger process \citep[e.g.,][]{yu11,Steinborn2015}, as well as being crucial to predicting the amplitudes and rates of GW signals that pulsar timing arrays and future space interferometers will detect \citep[e.g.,][]{Sesana2017}.

\subsection{Observational Evidence for Binary and Dual Supermassive Black Holes and Uncertainties}

Despite the theoretically appealing nature of this scenario, however, direct observational evidence for dual and binary SMBHs remains scarce. To date, only a single potential pc-scale binary SMBH is known,\footnote{There is one candidate sub-pc binary SMBH in OJ 287, which shows a $\sim$12 year quasi-periodic light curve, and is interpreted with a binary SMBH$+$accretion disk model \citep{sillanpaa88,valtonen08,Valtonen2016}. In addition, there is one candidate reported by \cite{boroson09}, which was later suggested to be an unusual disk emitter and not a binary SMBH \citep{chornock10}. There are $\sim$150 mili-pc binary SMBH candidates proposed based on quasi-periodic quasar light curves \citep[e.g.,][]{Graham2015a,Graham2015,Liutt2015,DOrazio2015a,DOrazio2016,Charisi2016,Zheng2015}, although the quasi-periodicity could also be due to single BH accretion disk instability and/or radio jet precession \citep[e.g.,][]{Kudryavtseva2011} or false periodicities caused by stochastic variability \citep[e.g.,][]{Vaughan2016}.} \objectname{B2~0402$+$379} \citep{maness04,rodriguez06}, with a (projected) separation of approximately 7 pc; one candidate sub-pc binary SMBH has just been identified in NGC 7674 from direct imaging \citep[with projected separation of $0.35$ pc;][]{Kharb2017} using Very long baseline interferometry (VLBI). The fraction of low-redshift AGN pairs on $\sim$5--100 kpc scales is a few percent \citep[e.g.,][]{Liu2011a,Liu2012}, and the fraction of intermediate-redshift binary quasars on tens to hundreds of kpc scales is $\sim 0.1$\% \citep[e.g.,][]{hennawi06,hennawi09,myers08,shen10c}. Until recently, on kpc scales, there had been only a handful of unambiguous cases of dual AGNs in which both SMBHs are detected in the radio \citep[e.g., \protect\objectname{3C~75},][]{owen85}, optical \citep[e.g., \protect\objectname{LBQS 0103-2753},][]{junkkarinen01}, or X-rays \citep[e.g., \protect\objectname{NGC~6240}, \protect\objectname{Mrk~463}, and \protect\objectname{Mrk~739};][]{komossa03,bianchi08,Koss2011}.

The advent of large and uniform spectroscopic surveys, such as the Sloan Digital Sky Survey \citep[SDSS;][]{York2000}, has enabled the identification of large numbers of galaxies with spectroscopic signatures potentially characteristic of dual AGNs on $\sim$kpc and sub-kpc scales \citep[e.g.,][]{wang09,Liu2010b,Smith2010,Ge2012,Barrows2013,Comerford2013,LyuLiu2016,Yuan2016}, as well as binary SMBHs on sub-pc scales \citep[e.g.,][]{tsalmantza11,eracleous11,Ju2013,shen13,Liu2014,Runnoe2015,Runnoe2017,Wang2017}. In particular, one such signature is having two spectral-line components associated with AGNs, such as \foiii , and with a velocity separation of a few hundred km s$^{-1}$ which signals orbital motion on galactic scales
\citep[e.g.,][]{zhou04,gerke07,comerford08,xu09,barrows12}, analogous to a double-lined spectroscopic star binary.

Higher angular resolution follow-up observations of candidates from systematic surveys have dramatically increased the number of dual AGNs on kpc-scale separations\citep[e.g.,][]{Liu2010a,Liu2017a,Fu2011a,Fu2012,mcgurk11,Shen2011,comerford11b}. However, there are considerable ambiguities associated with identifying dual SMBHs from spectral-line observations alone. Outflows associated with jets and rotating disks may also produce double-peaked narrow emission lines in AGNs \citep{axon98,xu09,crenshaw09,rosario10,Smith2010,smith11,comerford11a,fischer11,Shen2011}, and recent work has suggested that the majority ($\gtrsim 50\%$) of double-peaked narrow emission-line AGNs are likely due to complex narrow-line kinematics around single AGNs \citep{Shen2011,Fu2012,Nevin2016}.

\subsection{This Work: High-resolution Imaging with VLBA}

Considering the sizes of the narrow-line regions (NLRs) responsible for the \foiii\ lines and the evolutionary stages of a dual \hbox{SMBH}, the physical separation between dual AGNs with two distinct NLRs can be as small as $\sim$30 pc \citep[e.g.,][depending on AGN luminosity]{schmitt03,greene11}, corresponding to an angular separation of order 10 mas at a typical redshift $z \sim 0.1$; actual separations could be smaller when projected on the sky. Dual AGNs with such small separations would show double-peaked narrow emission lines given typical orbital velocities of the individual NLRs being a few hundred km s$^{-1}$ and typical velocity dispersions of $\lesssim$ a few hundred km s$^{-1}$ of the NLR gas clouds, provided that the two NLRs are not yet fully merged. Only with a full assessment of all the candidate dual AGNs at various separations will we be able to put robust constraints on aspects merger scenarios, such as the merger fraction, the dynamics of merging SMBH pairs, the significance of mergers in triggering AGNs, and the separations of the merging components at which AGNs are triggered. 

VLBI techniques routinely produce images with milliarcsecond resolutions, equivalent to a linear separation of order 10 pc (at $z \sim 0.1$), which is far higher than can be obtained at other wavelengths even with adaptive optics in the optical/near-infrared. Further, VLBI observations have traditionally been sensitive to high brightness temperature structures in the inner regions and nuclei of galaxies, such as AGNs and jets. Thus, the combination of spectroscopic surveys and VLBI imaging offers a powerful means of searching for sub-kpc dual SMBHs that is less biased due to spatial resolution limit compared with other observations. VLBI will also be sensitive to radio jets of scales of tens of parsecs; the gas outflows they drive may be responsible for the double-peaked narrow emission lines in AGNs.
In essence, VLBI observations can probe a new spatial regime in the parameter space of dual SMBHs or jets.

This paper presents Very Long Baseline Array \citep[VLBA;][]{Napier1994} observations of a subset of the double-peaked narrow emission-line AGNs identified by \cite{Liu2010b} from the SDSS DR7 \citep{SDSSDR7}.  The objective was to assess the radio detection rate and the fraction of objects that have compact and binary radio components in their cores. In~\S\ref{sec:observe}, we summarize our VLBA observations; in~\S\ref{sec:sources}, we discuss the six galaxies that we detect radio emission from; and in~\S\ref{sec:discuss}, we discuss what our results imply about sub-kpc dual SMBHs and implications for future observations. Throughout, we assume a cosmology of a Hubble constant of~70~km~s${}^{-1}$~Mpc${}^{-1}$, a matter density $\Omega_m = 0.3$, and a vacuum energy density of $\Omega_{\mathrm{vac}} = 0.7$.

\section{Target Selection, Observations, and Data Reduction and Analysis}\label{sec:observe}

As an initial search for close pairs (both physical and in projection) and to demonstrate the feasibility of a larger effort, we observed 13 type-2 Seyfert AGNs with double-peaked \OIIIc\ emission lines with VLBA (Program BL170, PI: Liu). The observations were at~3.6~cm (8.4~GHz) with a total bandwidth of~32~MHz.  This observational wavelength was motivated by two factors: (1) the angular resolution of the VLBA at~3.6~cm is approximately 1 mas, more than sufficient to detect any sub-kpc dual AGNs given their expected separations (\S\ref{sec:intro}) and (2) the VLBA is at its most sensitive at this wavelength.

\subsection{Target Selection}

\begin{figure}
 \begin{center}
  \includegraphics[width=0.48\textwidth]{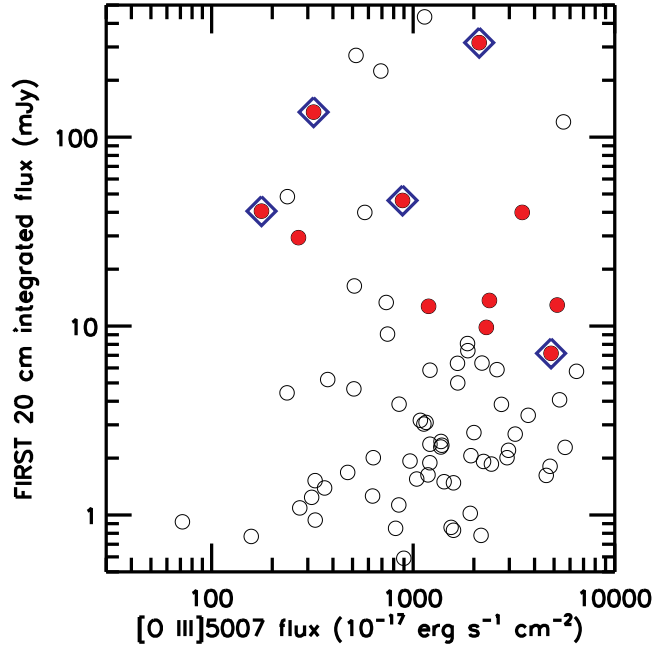}
  \caption{[O~{\tiny III}] emission-line flux versus FIRST integrated flux at 20 cm for our VLBA targets (red filled circles).
  Also shown for comparison are the other FIRST-detected sources (open circles) in the parent sample of double-peaked, narrow-line AGNs from \cite{Liu2010b}. 
  Typical measurement uncertainties are $\sim2.6\times10^{-16}$ erg s$^{-1}$ cm$^{-2}$ for [O~{\tiny III}] emission-line flux and $\sim$0.15 mJy for FIRST integrated flux at 20 cm.
  Targets that were detected by our VLBA observations are marked with blue open diamonds. 
  The VLBA-detected target \protect\objectname{SDSS~J135251.22$+$654113.2} is not shown here, because it was not covered by the FIRST survey.}
  \label{fig:target}
 \end{center}
\end{figure}

Our target sample was drawn from the 167 double-peaked emission-line AGNs identified by \cite{Liu2010b}. We first selected galaxies either having counterparts in the Faint Images of the Radio Sky at Twenty Centimeters survey \citep[\hbox{FIRST};][]{becker95} or that had existing observations within the VLA Archive from which the radio structure could be assessed. Of all the radio-bright objects in our parent sample (77 in 167 objects being detected by FIRST), we selected 12 objects that have 1.4 GHz flux densities above 7 mJy (from among a total of 23), and which appear unresolved\footnote{Had we have observed sources that were resolved by FIRST and not detected anything, one possibility would be that there is no milliarcsecond (mas) structure. Our preference toward unresolved FIRST sources is largely to assure a likely high detection rate.} in the core to \hbox{FIRST} (angular resolution $\sim$5''). The VLA Archival observations, where they exist, were used to confirm that the sources remain compact on sub-arcsec scales, to estimate their flux densities at higher frequencies to ensure adequate signal-to-noise ratios (S/N) for imaging, and improve the positional information for the new VLBA observations. We supplemented the sample with one additional object (\protect\objectname{SDSS~J135251.22$+$654113.2}) from the \cite{Liu2010b} sample that was not covered by the FIRST survey, but which had existing radio observations within the VLA Archive and satisfied our selection criteria. The median redshift of our target sample is $\sim$0.15. Figure \ref{fig:target} shows the [O~{\tiny III}] emission-line flux versus FIRST integrated flux at 20 cm for our targets as compared with the other radio-bright objects in our parent sample. Our VLBA targets are selected to have high radio fluxes at 20 cm, but their [O~{\tiny III}] emission-line fluxes sample the full range of the parent sample.

\begin{deluxetable*}{lcccccccc}
  \tablecaption{VLBA 3.6 cm (8.4 GHz) Detected Galaxies\label{tab:detect}}
  \tablewidth{\textwidth}
  \tabletypesize{\footnotesize}
  \tablehead{%
\colhead{} & 
\colhead{} & 
\colhead{$S_{{\rm FIRST}}$} &
\colhead{$\sigma_{{\rm FIRST}}$} &
\colhead{$\theta_{\mathrm{maj}} \times \theta_{\mathrm{min}}$} &
\colhead{$\sigma_I$} & 
\colhead{$I$} & 
\colhead{$S$} & 
\colhead{$L_\nu$} \\
\colhead{SDSS Name} &
\colhead{Redshift} & 
\colhead{(mJy)} &
\colhead{(\mjybm)} &
\colhead{(mas $\times$ mas)} & 
\colhead{(\mjybm)} & 
\colhead{(\mjybm)} & 
\colhead{(mJy)} & 
\colhead{($10^{23}$~W~Hz${}^{-1}$)} 
\\
\colhead{(1)} &
\colhead{(2)} & 
\colhead{(3)} & 
\colhead{(4)} & 
\colhead{(5)} & 
\colhead{(6)} & 
\colhead{(7)}  & 
\colhead{(8)}  & 
\colhead{(9)}  
        }
\startdata
\objectname{J091201.68$+$532036.6} & 0.1017 & 135.65 & 0.16 & 1.92 $\times$ 1.07 & 0.15 & 2.58 & 43.0
& 8.3 \\
\objectname{J113721.36$+$612001.2} & 0.1112 & 7.17 & 0.19 & 1.70 $\times$ 0.80 & 0.15 & 2.60 & 5.5 & 1.5 \\
\objectname{J124358.36$-$005845.4} & 0.4092 & 40.58 & 0.15 & 1.94 $\times$ 0.87 & 0.16 & 21.87 &
23.27 & 130 \\
\objectname{J135251.22$+$654113.2} & 0.2064 & N/A & N/A & 1.56 $\times$ 0.78 & 0.14 & 21.84 &
28.48 & 33 \\
\objectname{J231051.95$-$090011.9} & 0.0733 & 46.22 & 0.15 & 2.35 $\times$ 0.90 & 0.18 & 3.15 & 3.95
& 0.48 \\
\objectname{J233313.17$+$004911.8} & 0.1699 & 316.33 & 0.10 & 2.16 $\times$ 0.94 & 0.24 & 4.56 & 5.22
& 3.9 \\
\enddata
\tablecomments{
Column (1): SDSS designation with J2000 coordinates. 
Column (2): SDSS spectroscopic redshift. 
Column (3): FIRST integrated flux density at 20 cm; N/A means source is not covered by the FIRST survey.  
Column (4): RMS noise at 20 cm in the FIRST survey map; N/A means source is not covered by the FIRST survey.
Column (5): major and minor axis of the \textsc{clean} restoring beam. 
Column (6): 1-$\sigma$ noise level determined within a region 2$''$ square centered on the source.
Column (7): peak brightness.
Column (8): peak flux density.
Column (9): implied radio luminosity density.}
\end{deluxetable*}

\begin{deluxetable*}{lcccccc}
  \tablecaption{VLBA 3.6 cm (8.4 GHz) Undetected Galaxies\label{tab:undetect}}
  \tablewidth{\textwidth}
  \tablehead{%
\colhead{} & 
\colhead{} & 
\colhead{$S_{{\rm FIRST}}$} &
\colhead{$\sigma_{{\rm FIRST}}$} &
\colhead{$\theta_{\mathrm{maj}} \times \theta_{\mathrm{min}}$} & 
\colhead{$\sigma_I$} & 
\colhead{$L_\nu$} \\
\colhead{SDSS Name} & 
\colhead{Redshift} &
\colhead{(mJy)} &
\colhead{(\mjybm)} & 
\colhead{(mas $\times$ mas)} & 
\colhead{(\mjybm)} & 
\colhead{($10^{21}$~W~Hz${}^{-1}$)}\\
\colhead{(1)} & 
\colhead{(2)} & 
\colhead{(3)} & 
\colhead{(4)} & 
\colhead{(5)} & 
\colhead{(6)} & 
\colhead{(7)}
           }
\startdata
\objectname{J000911.58$-$003654.7} & 0.0733 &39.95 & 0.15 & 1.64 $\times$ 0.73 & 0.21 & $<$7.7 \\
\objectname{J073849.75$+$315611.9} & 0.2973 & 29.37 & 0.15 & 1.54 $\times$ 0.63 & 0.18 & $<$140  \\
\objectname{J080337.32$+$392633.1} & 0.0655 & 12.92 & 0.14 & 1.53 $\times$ 0.63 & 0.17 & $<$4.9 \\
\objectname{J085841.76$+$104122.1} & 0.1480 & 12.72 & 0.13 & 1.67 $\times$ 0.69 & 0.17 & $<$28  \\
\objectname{J110851.04$+$065901.4} & 0.1816 & 9.84 & 0.13 & 1.70 $\times$ 0.67 & 0.21 & $<$54 \\
\objectname{J135646.11$+$102609.1} & 0.1231 & 59.58 & 0.13 & 1.54 $\times$ 0.70 & 0.20 & $<$22 \\
\objectname{J171544.05$+$600835.7} & 0.1569 & 13.66 & 0.14 & 1.55 $\times$ 0.79 & 0.19 & $<$36  \\

\enddata
\tablecomments{
Column (1): SDSS designation with J2000 coordinates. 
Column (2): SDSS spectroscopic redshift. 
Column (3): FIRST integrated flux density at 20 cm.  
Column (4): RMS noise at 20 cm in the FIRST survey map.
Column (5): major and minor axis of the  \textsc{clean} restoring beam. 
Column (6): 1-$\sigma$ noise level determined within a region 2$''$ square centered on the source location.
Column (7): 3$\sigma$ upper limit on the radio luminosity density, at the galaxy's redshift.}
\end{deluxetable*}

\subsection{VLBA Observations}

The VLBA observations were conducted in four sessions, with each session lasting approximately 6 hr, in the interval between 2010 March 30 and June 5. Target flux densities at 8.4 GHz ranged from 135 to 1.5 mJy, with a median flux density of 40 mJy. Anticipating that some fraction of the flux density measured by FIRST might be resolved out by the VLBA observations, we used phase referencing for all of the observations, cycling between the target source (3~minutes) and a phase reference calibrator (2~minutes). Typical on-source integration times were~40~minutes, implying expected thermal noise levels of approximately 0.15 mJy beam$^{-1}$. The phase reference calibrator was selected from the VLBA calibrator database.  In addition, each session included short scans of a fringe finding calibrator (either \objectname{4C~39.25} or \objectname{3C~454.3}) and a source for checking the amplitude response (\objectname{J1310$+$3220}, \objectname{DA~193}, or \objectname{OJ~287}).

\subsection{Data Reduction and Analysis}

We adopted standard data reduction procedures using the Astronomical Image Processing System\footnote{http://www.aips.nrao.edu/index.shtml} (version 31DEC10). Specifically, we performed amplitude calibration of the visibility data using a priori knowledge of the system temperature information for the antennas, applied a parallactic correction for the rotation of the antenna feed orientation during the observation, determined the spectral bandpass response, and calculated residual delays and fringe rates (fringe fitting). A bandpass response function was determined because, even though the observations are intended to be continuum, the data were acquired in a spectral mode. These calibration steps were performed on the calibrators, most notably on the phase reference calibrators, and then interpolated onto the target sources. The full VLBA was used for these observations. Notionally, the array has 10 antennas, although because of the editing of data, the number of antennas being used in a given scan varied between nine and~10.

Each target source was then imaged. In all cases, a region approximately 2\arcsec\ $\times$ 2\arcsec\ was searched, corresponding to a linear distance of approximately 5~kpc at the median redshift of~0.15. Typical resolutions obtained were approximately 1.1 mas, corresponding to an equivalent linear distance of 2.7 pc, again at the median redshift. Tables~\ref{tab:detect} and~\ref{tab:undetect} summarize the characteristics of the images, and derived quantities from the images, for the detected and undetected galaxies, respectively. For all galaxies, we report the SDSS spectroscopic redshift~$z$; the major and minor axis of the \textsc{clean} restoring beam, $\theta_{\mathrm{maj}}$ and~$\theta_{\mathrm{min}}$; and the noise level in the image~$\sigma_I$. For the detected galaxies, we report the peak brightness~$I$ and the integrated flux density~$S$.  We also report the implied (spectral) luminosity~$L_\nu$ at the redshift of the detected galaxies and the (3$\sigma$) upper limit on~$L_\nu$ based on the image noise level for the undetected galaxies. For the sources that were detected, all were found within 0\farcs05 of their nominal position (determined either from optical images or from lower-resolution radio observations). For the sources that were not detected, the stated image noise levels were determined within the searched region.

\section{Individual Galaxies}\label{sec:sources}

In this section, we present the images of the detected galaxies and discuss our results in the context of other observations of these galaxies in the literature.  The images we present are much smaller than the full 2\arcsec\ $\times$ 2\arcsec\ region initially searched for emission as, in all cases, we found the emission to be not only compact but relatively close to the center of the region searched.

\begin{figure*}
 \begin{center}
  \includegraphics[width=0.54\textwidth]{J0912.eps}
  \includegraphics[width=0.6\textwidth]{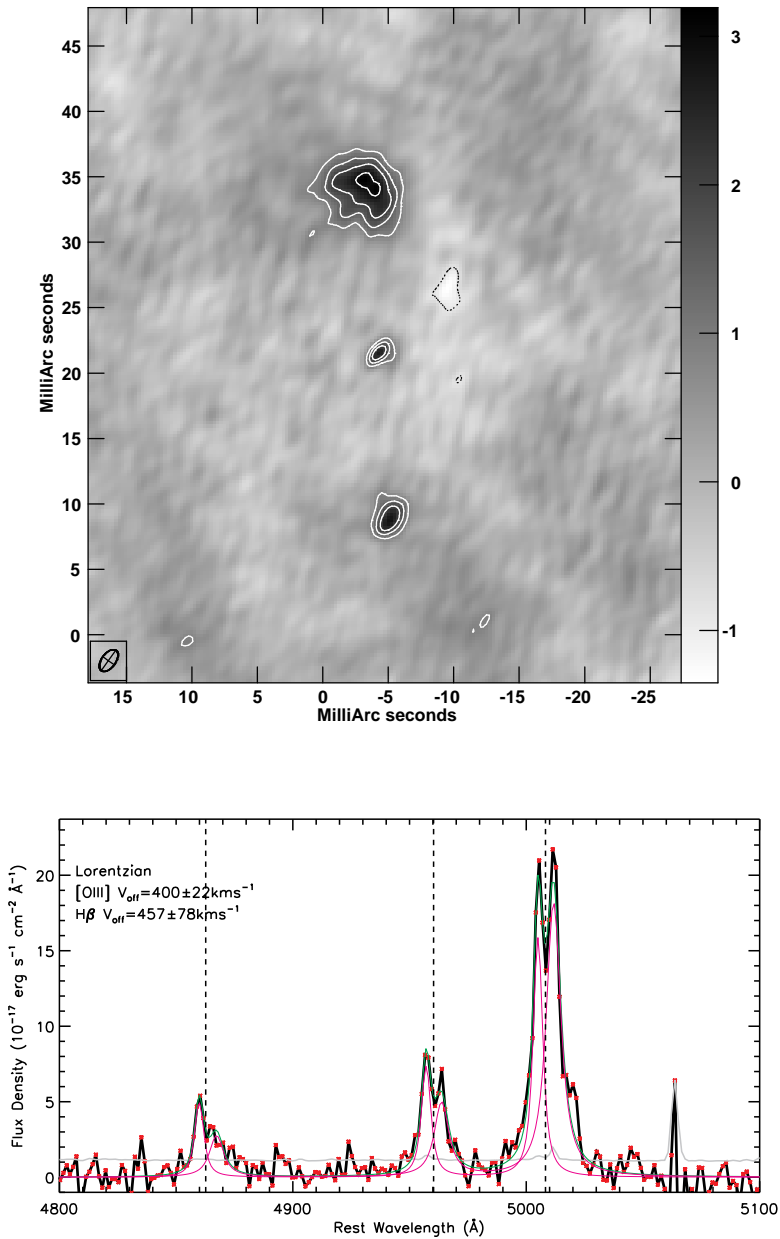}
  \caption{\footnotesize Top: VLBA 8.4 GHz image of \protect\objectname{SDSS~J091201.68$+$532036.6}.
    The gray scale is linear over the range $-1.34$ to 3.18 \mjybm.  The beam is 1.92~mas
    $\times$ 1.07 mas (corresponding to 3.6 pc $\times$ 2.0 pc at the redshift of the galaxy $z=0.1017$), and the noise level is 0.15 \mjybm.
    Contours are given with the levels set to be $-3$, 3, 5, 7, and~10 times the noise level in the image.
    Bottom: SDSS spectrum (flux density shown in red points connected with black curves and 1$\sigma$ error shown in gray; subtracted for host-galaxy stellar continuum) along with our best fits (model in green and individual velocity components in magenta) for the H$\beta$-[O {\tiny III}] region. The vertical lines are drawn at the systemic redshift from host-galaxy stellar absorption. Labeled on the plot are our best-fit model function (Lorentzian or Gaussian) and the velocity offsets between the double-peaked components measured for [O {\tiny III}]$\lambda\lambda$4959,5007 and for H$\beta$.
    }
  \label{fig:J091201.68+532036.6}
 \end{center}
\end{figure*}

\subsection{\protect\objectname{SDSS~J091201.68$+$532036.6}}\label{sec:J091201.68+532036.6}

The $z=0.1017$ AGN contains double-peaked \OIIIc\ lines in its SDSS fiber spectrum (Figure~\ref{fig:J091201.68+532036.6}), with peaks blueshifted and redshifted from the systemic velocity by 193 km s$^{-1}$ and 208 km s$^{-1}$, respectively \citep{Liu2010b}. SDSS images show that the host galaxy has a close companion to the southwest at a projected separation of 3.7 kpc ($2.''0$). It is unclear whether the SDSS fiber spectrum is significantly contaminated by light from the companion. We have a spectrum of the companion from MMT, but it shows no emission lines and does not have the S/N to measure a redshift. Its SDSS photometric redshift \citep{Oyaizu2008} is consistent with the redshift of the AGN.

\begin{figure*}
 \begin{center}
  \includegraphics[width=0.6\textwidth]{J1137.eps}
  \includegraphics[width=0.6\textwidth]{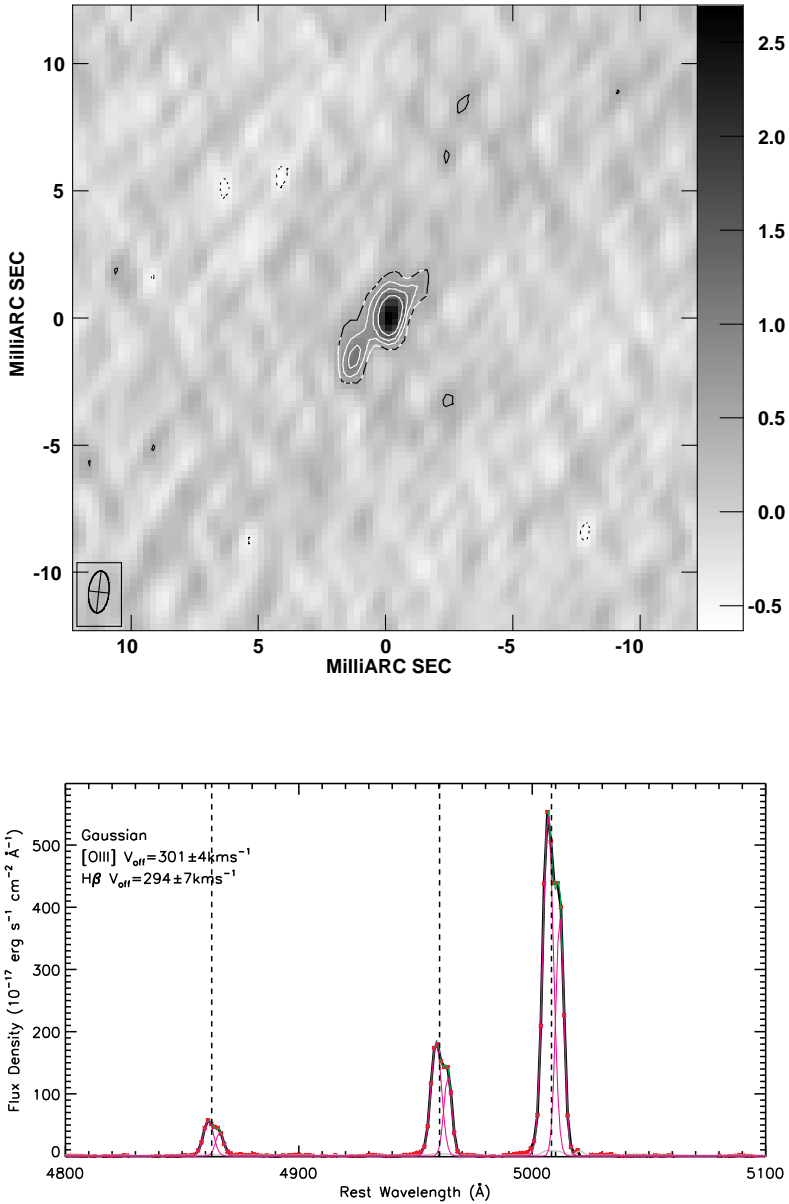}
  \caption{Top: VLBA 8.4 GHz image of \protect\objectname{SDSS~J113721.36$+$612001.2}.  The gray scale is linear over
  the range $-0.61$ to 2.67 \mjybm.  The beam is 1.70 mas
  $\times$ 0.80 mas (corresponding to 3.4 pc $\times$ 1.6 pc at the redshift of the galaxy $z=0.1112$), and the noise level is 0.15 \mjybm.
    Contours are given with the levels set to be $-3$, 3, 5, 7, and~10 times the noise level in the image.
  Bottom: SDSS spectrum (subtracted for host-galaxy stellar continuum) along with our best spectral fits for the H$\beta$-[O {\tiny III}] region. Figure captions and symbols are the same as those in the bottom panel of Figure \ref{fig:J091201.68+532036.6}.}
  \label{fig:J113721.36+612001.2}
 \end{center}
\end{figure*}

This galaxy contains a radio source that has been detected in surveys over the frequency range 0.365--8.4 GHz \citep{bwe91,gc91,dbbtw96,Condon1998}. In addition, Chandra X-ray observations show a point source with a flux of $6.8_{-1.4}^{+1.5} \times 10^{-15}$~erg~s${}^{-1}$~cm${}^{-2}$ \citep[0.5--7~\hbox{keV},][]{evans10}. In images from both the FIRST survey and the Cosmic Lens All-Sky Survey \citep[\hbox{CLASS},][]{Myers2003}, the radio source appears unresolved.  There is another, also unresolved, radio source approximately 30\arcsec\ to the southeast, but it is unclear whether the two are related.

Figure~\ref{fig:J091201.68+532036.6} shows the image resulting from our VLBA observations.  The radio source has three components, approximately aligned north--south.  The central component is unresolved, the northern component is clearly diffuse and extended, and the southern component is marginally resolved. The flux density obtained by CLASS is approximately 70 mJy whereas our VLBA observations recover approximately 43 mJy, indicating that there is (substantial) structure on sub-arcsecond scales that is resolved out by the VLBA observations.

In contrast to the other sources in this sample (Figures \ref{fig:J113721.36+612001.2}--\ref{fig:j233313.17+004911.8}, below), SDSS~J091201.68$+$532036.6 shows substantial sub-arcsecond--scale structure, though it also has a compact core component, with a flux density of 3 mJy, comparable to the flux densities measured for compact components of other sources in our sample.  While this source is unusual in our sample \citep[and that of][]{tingay11}, it is not unprecedented.  Multi-Element Radio-Linked Interferometer Network (MERLIN) observations of sources commonly show structure on sub-arcsecond scales \citep[e.g.,][]{Filho2006,Williams2017}, most notably including the double-peaked emission-line source \objectname{3C~316} \citep{An2013}.

Moreover, we note two potential selection effects. First, while the extended emission contributes significantly to the total flux density of the source, its typical surface brightness is a factor of 2--3 lower than that of the compact component or the peak brightnesses in the north and south components.  Modest changes in the observations
(e.g., had we observed for only 20 minutes instead of 40 minutes) would have resulted in less of the extended emission being apparent. Second, typical observations of lower luminosity radio sources have sought to address whether there is a compact core component. With such a focus, it is possible that diffuse emission located tens of
milliarcseconds from a core component would not have been noticed. Indeed, a limited sampling of the literature shows a number of VLBI images of lower luminosity radio sources for which the typical image is 10~mas $\times$ 10~mas (cf. Figure~\ref{fig:J091201.68+532036.6}).

\subsection{\protect\objectname{SDSS~J110851.04$+$065901.4}}\label{J110851.04+065901.4}

This $z=0.1816$ AGN contains double-peaked \OIIIc\ lines in its SDSS fiber spectrum, with peaks blueshifted and redshifted from the systemic velocity by 95 km s$^{-1}$ and 114 km s$^{-1}$, respectively \citep{Liu2010b}.  It represents our only target that hosts a kpc-scale dual AGN detected by Chandra in the X-rays \citep{Liu2013}. 

\cite{Liu2010a} resolved a dual AGN in this galaxy using near-IR imaging and optical slit spectroscopy, with an approximate separation of 0\farcs7, corresponding to a linear separation of~2.1~kpc at its redshift \citep{Liu2013}. Both ground-based imaging in the NIR \citep{Liu2010a,Fu2012} and HST $Y$-band imaging \citep{Liu2013} show disturbance in the galaxy surface brightness profile, suggesting tidal interactions. \citet{Liu2013} confirmed its dual AGN nature in the X-rays based on Chandra observations combined with constraints on X-ray contribution from star formation estimated from HST $U$-band imaging. We do not detect a radio source, at a 5$\sigma$ limit of~1.1~\mjybm.

\subsection{\protect\objectname{SDSS~J113721.36$+$612001.2} (\protect\objectname{4C~61.23})}\label{J113721.36+612001.2}

This $z=0.1112$ AGN contains double-peaked \OIIIc\ lines in its SDSS fiber spectrum (Figure~\ref{fig:J113721.36+612001.2}), with peaks blueshifted and redshifted from the systemic velocity by 87 km s$^{-1}$ and 214 km s$^{-1}$, respectively \citep{Liu2010b}. The SDSS images show no evidence for tidal disturbance, which is also confirmed by an {\it HST} $Y$-band image (in preparation). This galaxy contains a radio source that has been detected in surveys at least over the frequency range 38 MHz to 5 GHz \citep{gsw67,bwe91,gc91,White1992,hwrw95,dbbtw96,Cohen2007}. \cite{lcfgmmv01} classify it as a Faranoff-Riley~II radio galaxy.  Their VLA image (at~5~GHz) shows it to be symmetric, with the lobes oriented at~135\arcdeg\ E from N.

Figure~\ref{fig:J113721.36+612001.2} shows our VLBA image. Although superficially an apparent double, with the second component approximately 2~mas to the southeast of the primary component, the orientation of the milliarcsecond structure is similar to that of the arcsecond FR~II structure, suggesting that the small-scale fainter component could be a portion of the jet that is feeding the southeast large-scale radio lobe.

\begin{figure*}
 \begin{center}
  \includegraphics[width=0.6\textwidth]{J1243.eps}
  \includegraphics[width=0.6\textwidth]{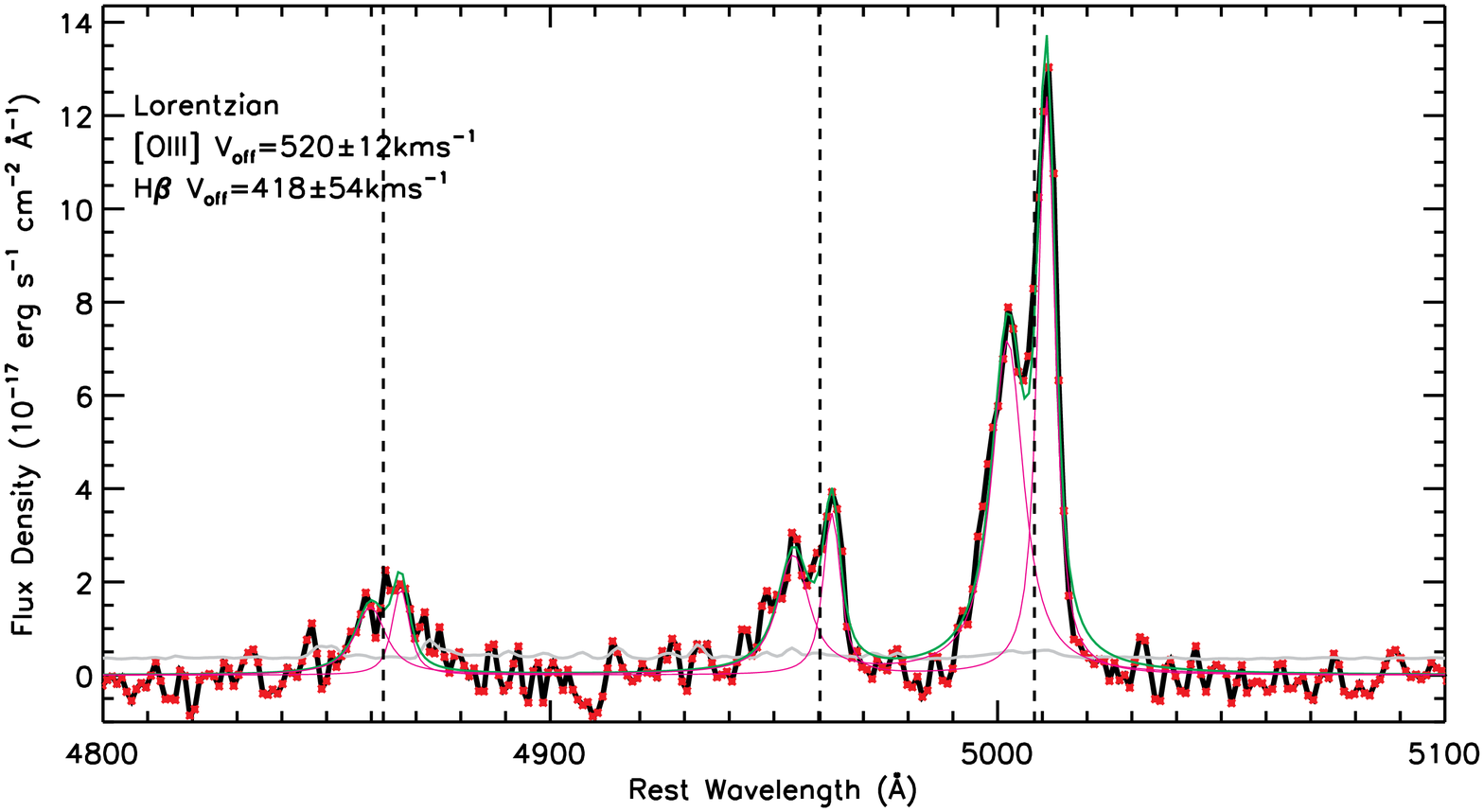}
  \caption{Top: VLBA 8.4 GHz image of \protect\objectname{SDSS~J124358.36$-$005845.4}.  The gray
  scale is logarithmic, with a maximum at~22.08~\mjybm.
  The beam is 1.94 mas $\times$ 0.87 mas (corresponding to 11 pc $\times$ 4.7 pc at the redshift of the galaxy $z=0.4092$), and the noise level is 0.16 \mjybm.
    Contours are given with the levels set to be $-3$, 3, 5, 7, and 10 times the noise level in the image.
  Bottom: SDSS spectrum (subtracted for host-galaxy stellar continuum) along with our best spectral fits for the H$\beta$-[O {\tiny III}] region. Figure captions and symbols are the same as those in the bottom panel of Figure \ref{fig:J091201.68+532036.6}.
  }
  \label{fig:J124358.36-005845.4}
 \end{center}
\end{figure*}

\subsection{\protect\objectname{SDSS~J124358.36$-$005845.4}}\label{sec:J124358.36-005845.4}

This $z=0.4092$ AGN shows double-peaked \OIIIc\ lines in its SDSS fiber spectrum (Figure~\ref{fig:J124358.36-005845.4}), with peaks blueshifted and redshifted from the systemic velocity by 360 km s$^{-1}$ and 161 km s$^{-1}$, respectively \citep{Liu2010b}. The SDSS images show no evidence for tidal disturbance, although both the image quality and sensitivity may be too low to put a strong constraint. This galaxy contains a radio source that has been detected in surveys over the frequency range 1.4--5 GHz \citep{gwbe95,Condon1998}, and the radio source appears unresolved in \hbox{FIRST}. Figure~\ref{fig:J124358.36-005845.4} shows that the radio source has a compact, unresolved component in our VLBA image. The flux density on these scales is approximately 23 mJy at 3.6 cm, whereas the FIRST flux density is approximately 40 mJy (at 20 cm).  The spectrum of the radio source is approximately flat, suggesting that nearly 50\% of the flux density is not recovered by the VLBA image.

Within a region of size $\pm 1\arcsec$, there are no other radio sources stronger than 1.15 mJy (7.5$\sigma$).  The nearly equatorial declination and limited hour angle coverage contribute to a point spread function (beam) with high secondary peaks, such that a more stringent limit cannot be placed.

\subsection{\protect\objectname{SDSS~J135251.22$+$654113.2}}\label{sec:J135251.22+654113.2}

\begin{figure*}[tb]
 \begin{center}
  \includegraphics[width=0.6\textwidth]{J1352.eps}
  \includegraphics[width=0.6\textwidth]{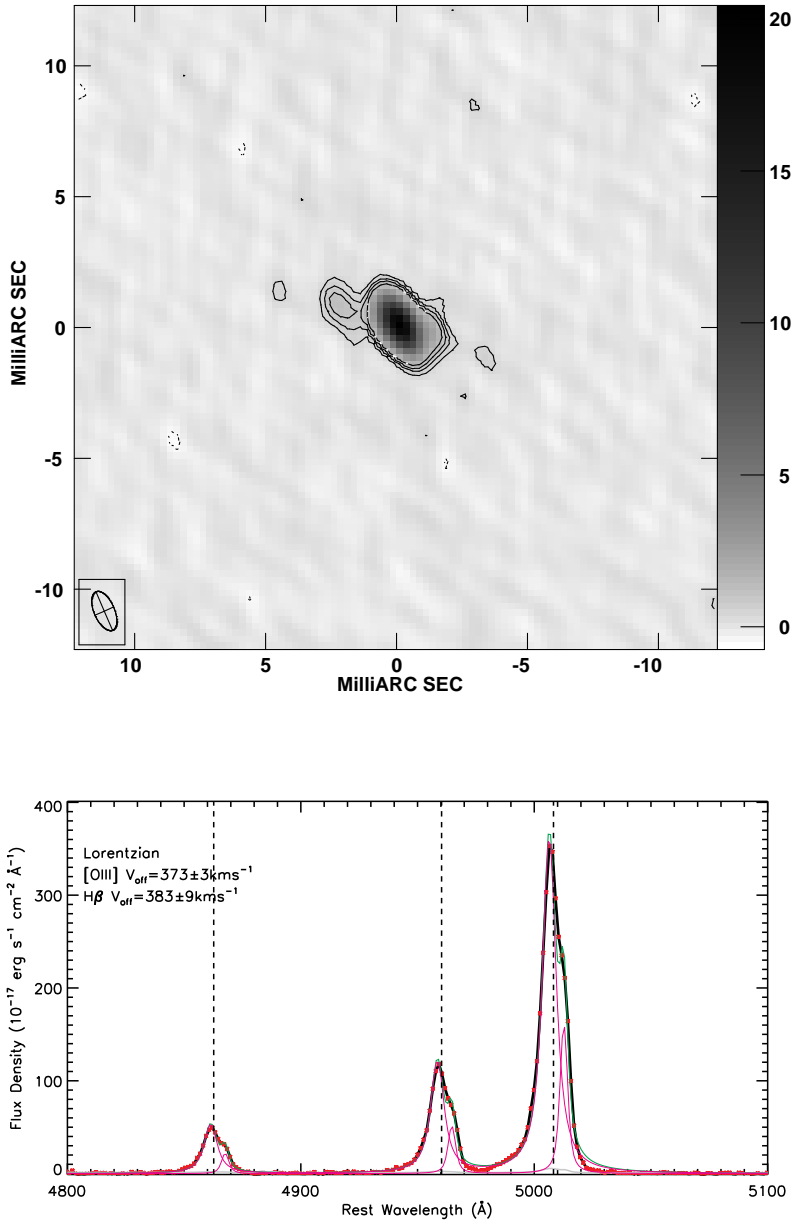}
  \caption{Top: VLBA 8.4 GHz image of \protect\objectname{SDSS~J135251.22$+$654113.2}.  The gray scale is
  logarithmic, with a maximum at 20.29 \mjybm.  The beam is 1.56 mas
  $\times$ 0.78 mas (corresponding to 5.3 pc $\times$ 2.6 pc at the redshift of the galaxy $z=0.2064$), and the noise level is 0.14 \mjybm.
    Contours are given with the levels set to be $-3$, 3, 5, 7, and 10 times the noise level in the image.
  Bottom: SDSS spectrum (subtracted for host-galaxy stellar continuum) along with our best spectral fits for the H$\beta$-[O {\tiny III}] region. Figure captions and symbols are the same as those in the bottom panel of Figure \ref{fig:J091201.68+532036.6}.
  }
  \label{fig:J135251.22+654113.2}
 \end{center}
\end{figure*}

This $z=0.2064$ AGN shows double-peaked \OIIIc\ lines in its SDSS fiber spectrum (Figure~\ref{fig:J135251.22+654113.2}), with peaks blueshifted and redshifted from the systemic velocity by 108 km s$^{-1}$ and 265 km s$^{-1}$, respectively \citep{Liu2010b}. The SDSS images show tentative evidence for tidal disturbance, possibly related to a companion to the southeast of the main galaxy. 

This galaxy contains a radio source which has been detected in surveys at least over the frequency range 74 MHz to 8 GHz \citep{hmwb90,bwe91,gc91,White1992,dbbtw96,Condon1998,Cohen2007}.  The radio source is unresolved in the CLASS image, with a flux density of 79 mJy.

Figure~\ref{fig:J135251.22+654113.2} shows our VLBA image. There is at least one faint component located approximately 2.4 mas to the east of the primary component; there may also be a fainter component farther to the east. Our VLBA observations recover approximately 30 mJy, indicating that there is likely to be substantial structure on sub-arcsecond scales. 

Within a region of size $\pm 1\arcsec$, there are no other apparent radio sources brighter than 0.79~\mjybm\ (6.6$\sigma$). Strictly, this is brighter than the nominal statistical threshold, but the brightest pixels in the residual image appear near the edges of the image and there are (negative) pixels with comparable absolute brightnesses.

\subsection{\protect\objectname{SDSS~J231051.95$-$090011.9}}\label{sec:J231051.95-090011.9}

\begin{figure*}
 \begin{center}
  \includegraphics[width=0.6\textwidth]{J2310.eps}
  \includegraphics[width=0.6\textwidth]{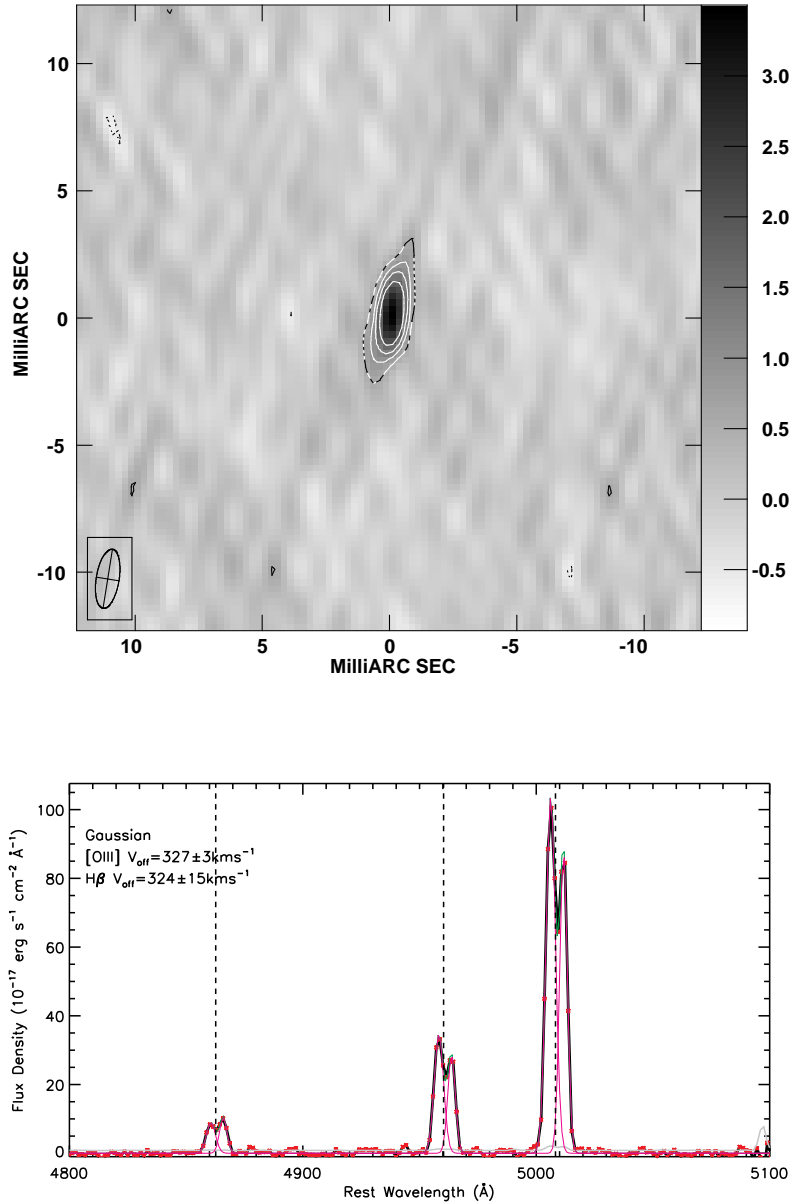}
  \caption{Top: VLBA 8.4 GHz image of \protect\objectname{SDSS~J231051.95$-$090011.9}.  The gray
    scale is linear over the range $-0.90$ to 3.47 \mjybm.
    The beam is 2.35 mas $\times$ 0.90 mas (corresponding to 4.1 pc $\times$ 1.6 pc at the redshift of the galaxy $z=0.0944$), and the noise level is 0.18 \mjybm.
    Contours are given with the levels set to be $-3$, 3, 5, 7, and 10 times the noise level in the image.
    Bottom: SDSS spectrum (subtracted for host-galaxy stellar continuum) along with our best spectral fits for the H$\beta$-[O {\tiny III}] region. Figure captions and symbols are the same as those in the bottom panel of Figure \ref{fig:J091201.68+532036.6}.
    }
  \label{fig:j231051.95-090011.9}
 \end{center}
\end{figure*}

This $z=0.0944$ AGN shows double-peaked \OIIIc\ lines in its SDSS fiber spectrum (Figure~\ref{fig:j231051.95-090011.9}), with peaks blueshifted and redshifted from the systemic velocity by 121 km s$^{-1}$ and 206 km s$^{-1}$, respectively \citep{Liu2010b}. The SDSS images show no evidence for a double stellar core, which is further confirmed by $K_s$-band imaging with Magellan/PANIC \citep{Shen2011} at $0.''6$ resolution and by HST ACS/F606W imaging at $0.''1$ resolution \citep{Fu2012}. However, there is tentative evidence for tidal disturbance, both from the SDSS images and the $K_s$-band images presented by \citet{Shen2011}. There is an apparent companion $\sim4''$ away to the southwest at PA=73$^{\circ}$, which does not contribute to the \foiii\ emission observed in the SDSS fiber spectrum. \citet{Shen2011} presented a slit spectrum from the Apache Point Observatory 3.5m Dual Imaging Spectrograph at PA=73$^{\circ}$, which suggested that the two velocity components in \foiii\ were spatially unresolved at $2''$ resolution.

The FIRST image shows this galaxy to be dominated by a compact component, with what may be a faint jet extending to the southeast. Figure~\ref{fig:j231051.95-090011.9} shows our VLBA image to consist of a single component, which is either unresolved or just marginally resolved.  The orientation of the milliarcsecond-scale structure may appear consistent with that in the FIRST image, but this orientation is essentially the same as the \textsc{clean} beam (which is an elliptical gaussian with a position angle of~$-9\arcdeg$, east of north).

\subsection{\protect\objectname{SDSS~J233313.17$+$004911.8} (\protect\objectname{PKS~2330$+$005})}\label{sec:J233313.17+004911.8}

\begin{figure*}
 \begin{center}
  \includegraphics[width=0.6\textwidth]{J2333.eps}
  \includegraphics[width=0.6\textwidth]{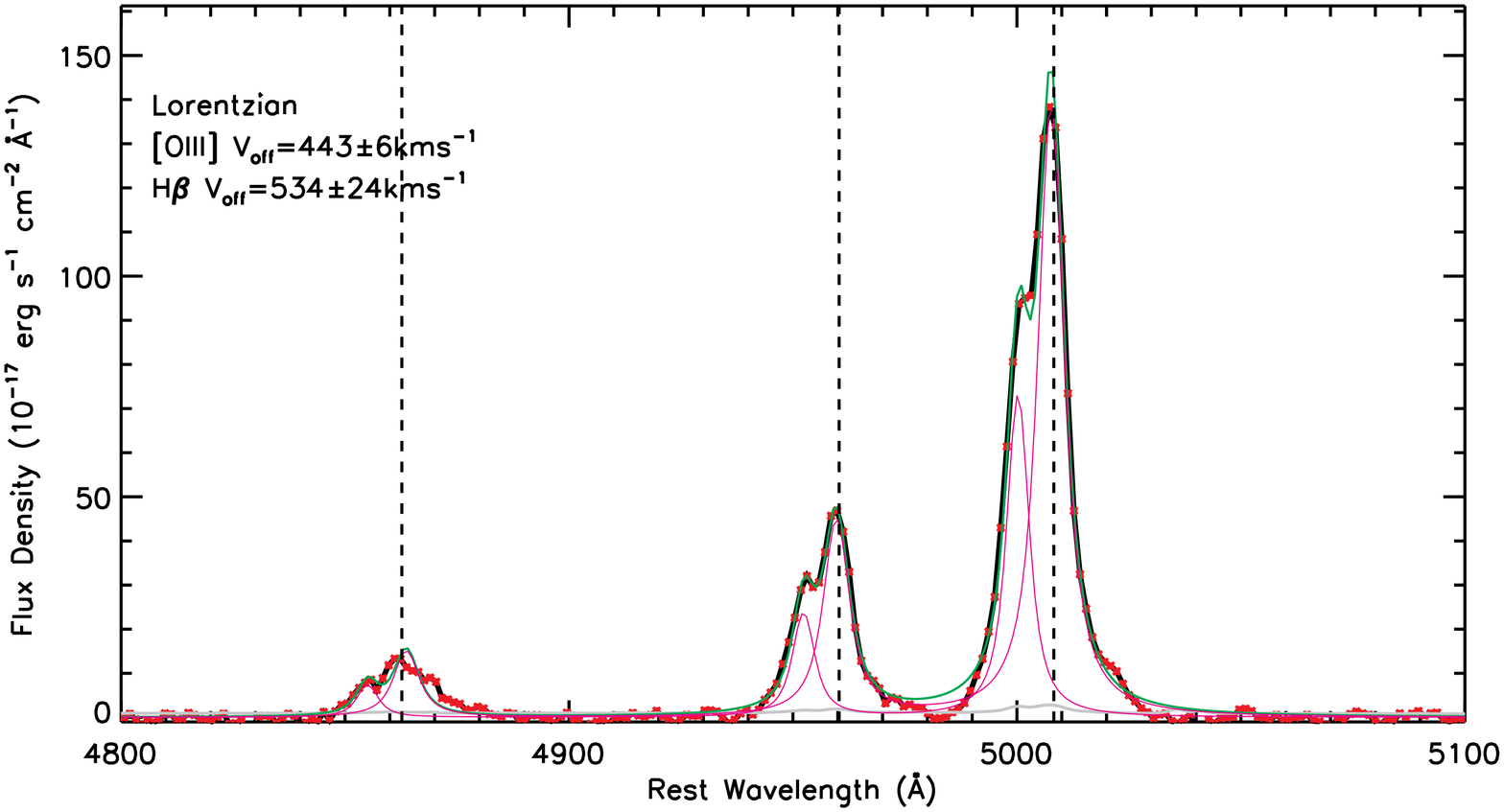}
  \caption{Top: VLBA 8.4 GHz image of \protect\objectname{SDSS~J233313.17$+$004911.8}.  The gray
    scale is linear over the range $-1.43$ to 4.74 \mjybm.
    The beam is 2.16 mas $\times$ 0.94 mas (corresponding to 6.3 pc $\times$ 2.7 pc at the redshift of the galaxy $z=0.1699$), and the noise level is 0.24 \mjybm.
    Contours are given with the levels set to be $-3$, 3, 5, 7, and 10 times the noise level in the image.
    Bottom: SDSS spectrum (subtracted for host-galaxy stellar continuum) along with our best spectral fits for the H$\beta$-[O {\tiny III}] region. Figure captions and symbols are the same as those in the bottom panel of Figure \ref{fig:J091201.68+532036.6}.
    }
  \label{fig:j233313.17+004911.8}
 \end{center}
\end{figure*}

This $z=0.1699$ AGN shows double-peaked \OIIIc\ lines in its SDSS fiber spectrum (Figure~\ref{fig:j233313.17+004911.8}), with both peaks blueshifted from the systemic velocity by 480 km s$^{-1}$ and 36 km s$^{-1}$, respectively \citep{Liu2010b}. The SDSS images show no evidence for a double stellar core, which is confirmed by $K_s$-band imaging with Magellan/PANIC \citep{Shen2011} at $0.''8$ resolution and by $K_p$-band imaging with Keck II/NIRC2 LGSAO at $0.''1$ resolution \citep{Fu2012}. However, there is tentative evidence for tidal disturbance, both from SDSS images and the $K_s$-band images presented by \citet{Shen2011}. There is an apparent companion $\sim6''$ away to the southeast at PA=117$^{\circ}$, which does not contribute to the \foiii\ emission observed in the SDSS fiber spectrum. \citet{Shen2011} presented a slit spectrum from the APO 3.5m DIS at PA=117$^{\circ}$, which suggested that the two velocity components in \foiii\ were spatially unresolved at $1.''5$ resolution.

This galaxy contains a radio source that has been detected in surveys from 74 MHz to 5 GHz \citep{blbhm86,Wright1990,bwe91,gc91,White1992,gwbe95,dbbtw96,Condon1998,Cohen2007}. We are unaware of any previous published observations in which the radio source is resolved. Figure~\ref{fig:j233313.17+004911.8} shows our VLBA image to consist of a single component.

\section{Results and Discussion}\label{sec:discuss}

\subsection{Detection Rate}

Of our initial sample of~13 SDSS galaxies showing double-peaked \foiii\ emission lines, and with either FIRST or previous VLA detections, we detected six (46\%) in our VLBA observations to a flux density of $\sim$1 mJy. Of these six, two show likely jet structures on sub-kpc scales, while the other four are unresolved. In the other seven objects without a detectable radio core, most of the radio emission is likely from larger-scale jets or lobes. These sources are all sufficiently radio luminous that the radio emission is unlikely to be dominated by star formation in the host galaxies. The lack of mas-scale structure could also indicate that the central engine has recently shut down, although this explanation would require fine tuning, i.e., the fuel to the central engine shuts down only so recently that we do not see the central engine but we still see some lobe emission. As shown in Figure \ref{fig:target}, the VLBA-detected targets on average have higher integrated fluxes at 20 cm from FIRST than the VLBA non-detected sources. On the other hand, the [O~{\tiny III}] emission-line fluxes for the VLBA-detected and non-detected samples are similar. In no case have we detected an unambiguous sub-kpc dual AGN.

\subsection{Implications on the Population of Sub-kpc Dual AGNs}

We now consider what our observations imply about the potential population of sub-kpc dual AGNs. The fraction of dual AGNs on kpc scales among double-peaked emission-line AGN is estimated to be of order 10\% \citep[e.g.,][]{Shen2011,Fu2012}. Another $\sim 40$\% of objects in the \citet{Shen2011} sample are ambiguous and some of them could harbor a sub-kpc scale dual AGN that is unresolvable with optical/NIR observations. However, we detected no sub-kpc scale dual AGN out of 13 objects in our VLBA observations, which may hint at a similar fraction ($\lesssim 10$\%) of sub-kpc dual AGNs among these double-peaked \foiii\ AGN. This would strengthen our conclusion in \citet{Shen2011} that most double-peaked profiles are caused by NLR kinematics in single AGNs rather than by the orbital motion of dual AGNs.

Another possibility is that some of the galaxies that we detect do contain a dual AGN, but that the second SMBH in the system is either radio faint or is of sufficiently low mass as to be undetectable in our radio observations.  Accreting black holes display a ``fundamental plane'' relationship between their X-ray luminosity, radio luminosity, and mass \citep{Merloni2003,Falcke04,gkmdmr09}, which allows us to estimate a BH mass given measurements in radio and X-rays.  Accordingly, we have searched the Chandra X-ray archives for available observations of these galaxies.  Only \objectname{SDSS~J091201.68$+$532036.6} (\S\ref{sec:J091201.68+532036.6}) and \objectname{SDSS~J110851.04$+$065901.4} \citep{Liu2013} have been observed and detected previously in targeted X-ray observations. For the remaining galaxies, we examined the ROSAT All-Sky Survey\footnote{http://www.xray.mpe.mpg.de/cgi-bin/rosat/data-browser} for a possible X-ray counterpart.  In no case, do we detect a ROSAT X-ray source at the location of these galaxies. We converted the ROSAT X-ray flux limits to rest-frame ~2--10~keV flux limits using the online PIMMS tool\footnote{http://cxc.harvard.edu/toolkit/pimms.jsp}.

\begin{deluxetable}{lccc}
\tablecaption{Radio and X-Ray Luminosities, and Estimates on Black Hole Masses\label{tab:bh}}
\tablewidth{0pc}
\tablehead{%
 \colhead{} & 
 \colhead{$L_R$} & 
 \colhead{$L_X$} & 
 \colhead{$M_{\mathrm{BH}}$} \\
 \colhead{SDSS Name} & 
 \colhead{($10^{38}$~erg~s${}^{-1}$)} & 
 \colhead{($10^{40}$~erg~s${}^{-1}$)} & 
 \colhead{($10^8$~$M_\sun$)}  \\
 \colhead{(1)} & 
 \colhead{(2)} & 
 \colhead{(3)} & 
 \colhead{(4)}
 }
\startdata
\objectname{J091201.68$+$532036.6} & 34.0  & $8.3_{-1.7}^{+1.9}$ & 5.1 $\pm$ 0.3 \\
                                        & $<$23.8  & $8.3_{-1.7}^{+1.9}$ & $<$4.3 \\
\objectname{J113721.36+612001.2} &  10.92 &   $<$4.02 & $<$3.5 \\
\objectname{J124358.36-005845.4} & 225.12 & $<$101 & $<$6.9 \\
\objectname{J135251.22+654113.2} &  34.44 &   $<$9.05 & $<$5.0 \\
\objectname{J231051.95-090011.9} &   9.24 &   $<$6.26 & $<$2.9 \\
\objectname{J233313.17+004911.8} &  45.36 &  $<$25.5 & $<$4.4 \\

\\

\objectname{J000911.58-003654.7} &   $<$6.47 & $<$3.52 & $<$2.8 \\
\objectname{J073849.75+315611.9} & $<$118 &  $<$73.7 & $<$5.4 \\
\objectname{J080337.32+392633.1} &   $<$4.12 &   $<$3.25 & $<$2.3 \\
\objectname{J085841.76+104122.1} &  $<$23.5 &  $<$19.2 & $<$3.5 \\
\objectname{J110851.04+065901.4} &  $<$45.4 &  $<$22.7 & $<$4.6 \\
\objectname{J135646.11+102609.1} &  $<$18.5 &  $<$12.0 & $<$3.5 \\

\enddata
\tablecomments{
Galaxies are divided into two sets according to whether they are
detected or not by VLBA.  The top set corresponds to galaxies detected in our
VLBA observations (see also Table~\ref{tab:detect}), the lower set to the undetected
galaxies (see also Table~\ref{tab:undetect}).
Column (1): SDSS designation with J2000 coordinates. 
Column (2): Radio luminosity or 3-$\sigma$ upper limit. 
Column (3): X-ray luminosity or 3-$\sigma$ upper limit. 
Column (4): Estimated black hole mass or upper limit assuming the fundamental plane relationship between X-ray luminosity, radio luminosity and black hole mass \citep{gkmdmr09}.
}
\end{deluxetable}

Table~\ref{tab:bh} summarizes what the combined radio and X-ray measurements imply about the masses, or upper limits on the masses, of SMBHs in the nuclei of these galaxies. For the galaxy with both radio and X-ray detections (\objectname{SDSS~J091201.68$+$532036.6}), we estimate the masses of a potential dual SMBH system (one being radio bright and one being radio faint, i.e., below our detection limit) making the following assumptions. For the radio factor in the fundamental plane relation, we use either the flux density of the compact component or the 3$\sigma$ upper limit. For the X-ray factor in the fundamental plane relation, we assume that the two putative SMBHs would contribute equally to the X-ray flux. While this approach clearly does not yield a unique solution, it is indicative of the characteristics of a dual SMBH system, if the second black hole is not radio bright.

For the remaining galaxies, we use the (3$\sigma$) upper limit in the radio image and that on the X-ray flux to constrain the mass of an SMBH in the galaxy\footnote{For consistency with the other radio undetected galaxies, we also quote the upper limit on the X-ray flux for the galaxy \objectname{SDSS~J110851.04$+$065901.4}, which is detected by Chandra.}.  This approach is also clearly not unique, but it should suffice to provide an estimate of the possible SMBH masses.

Typical 3$\sigma$ upper limits on the mass of a second SMBH in the nuclei of these galaxies are (3--7) $\times 10^8$ $M_\sun$.  We stress that these black hole mass limits should be viewed as indicative, particularly given that \cite{gkmdmr09} find that the fundamental plane relation that we have assumed has a scatter of~0.77~dex. Nonetheless, it is possible that these galaxies might still contain a second SMBH, whose presence may become apparent with significantly deeper radio and X-ray observations.

\subsection{Comparison with Previous Work and Remarks on Future Directions}

\cite{tingay11} have conducted a similar search for dual SMBHs with VLBA. While their fraction of detected sources was lower (2 of~12 or 17\%), they also found no dual AGN candidates.  Taken together, these observations are consistent with the fraction of (radio-bright) dual SMBHs on sub-kpc to pc-scale separations being similar to that on larger separations, namely 0.1\%, though the sample remains small. As nearly half of the objects in the \cite{Liu2010b} sample have a radio counterpart and that new samples of AGNs with double-peaked narrow emission lines are available \citep[e.g.,][]{LyuLiu2016,Yuan2016}, there are ample possibilities for expanding the set of double-peaked line emission galaxies that have been examined for possible dual AGNs on sub-kpc scales.

Having a higher sensitivity than VLBA on larger scales, VLA is a better match in terms of searching for kpc-scale dual AGNs \citep[e.g.,][]{Burke-Spolaor2014,Fu2015,Muller-Sanchez2015}, kpc-scale jet-cloud interactions, or kpc-separation compact radio sources. Nevertheless, VLBA is superior in terms of spatial resolution, and has the potential to resolve sub-kpc projected pairs, and to test the probability of sub-kpc jets as the origin for double-peaked narrow-line profiles, where the NLR emission is unresolved on kpc scales \citep[e.g.,][]{Wrobel2014a}. It would also be interesting to carry out radio follow ups for new large samples of AGNs with double-peaked narrow emission lines at higher redshift \citep[e.g.,][]{LyuLiu2016,Yuan2016} to address their possible redshift and/or luminosity evolution \citep[e.g.,][]{yu11}.

\acknowledgements

We thank S.~Burke-Spolaor for helpful discussions and the anonymous referee for a careful and useful report that improved the paper. 
Y.S. acknowledges support from the Alfred P. Sloan Foundation and NSF grant 1715579. 

The NANOGrav project receives support from National Science Foundation Physics Frontier Center award number 1430284. The Long Baseline Observatory is a facility of the National Science Foundation operated under cooperative agreement by Associated Universities, Inc. This research has made use of the NASA/IPAC Extragalactic Database (NED) which is operated by the Jet Propulsion Laboratory, California Institute of Technology, under contract with the National Aeronautics and Space Administration. This research has made use of NASA's Astrophysics Data System. This research has made use of data obtained from the High Energy Astrophysics Science Archive Research Center (HEASARC), provided by NASA's Goddard Space Flight Center. This research has made use of data obtained from the Chandra Source Catalog, provided by the Chandra X-ray Center (CXC) as part of the Chandra Data Archive. Part of this research was carried out at the Jet Propulsion Laboratory, California Institute of Technology, under a contract with the National Aeronautics and Space Administration.

Funding for the SDSS and SDSS-II has been provided by the Alfred P. Sloan Foundation, the Participating Institutions, the National Science Foundation, the U.S. Department of Energy, the National Aeronautics and Space Administration, the Japanese Monbukagakusho, the Max Planck Society, and the Higher Education Funding Council for England. The SDSS website is http://www.sdss.org/.

The SDSS is managed by the Astrophysical Research Consortium for the Participating Institutions. The Participating Institutions are the American Museum of Natural History, Astrophysical Institute Potsdam, University of Basel, University of Cambridge, Case Western Reserve University, University of Chicago, Drexel University, Fermilab, the Institute for Advanced Study, the Japan Participation Group, Johns Hopkins University, the Joint Institute for Nuclear Astrophysics, the Kavli Institute for Particle Astrophysics and Cosmology, the Korean Scientist Group, the Chinese Academy of Sciences (LAMOST), Los Alamos National Laboratory, the Max Planck-Institute for Astronomy (MPIA), the Max Planck Institute for Astrophysics (MPA), New Mexico State University, Ohio State University, University of Pittsburgh, University of Portsmouth, Princeton University, the United States Naval Observatory, and the University of Washington.

\textit{Facilities:} \facility{VLBA}, Sloan

\bibliography{/Users/zeus/Documents/References/binaryrefs}

\end{document}